\documentclass{webofc}
\usepackage[varg]{txfonts}   
\usepackage[rightcaption,raggedright]{sidecap}
\usepackage{ragged2e}
\begin{document}
\title{Bayesian calibration of viscous anisotropic hydrodynamic (VAH) simulations of heavy-ion collisions}
%
%

\author{\firstname{Ulrich} \lastname{Heinz}\inst{1}\fnsep\thanks{\email{heinz.9@osu.edu}} \and
        \firstname{Dananjaya} \lastname{Liyanage}\inst{1}\fnsep\thanks{\email{liyanagedananjaya@gmail.com}} \and
        \firstname{Cullen} \lastname{Gantenberg}\inst{1,2}\fnsep\thanks{\email{cgantenb@uw.edu
        }}
}

\institute{Physics Department, The Ohio State University, Columbus, OH 43210-1117, USA 
\and
          Department of Physics, University of Washington, Seattle WA 98195-1560, USA 
          }

\abstract{%
A Bayesian calibration, using experimental data from 2.76\,$A$\,TeV Pb-Pb collisions at the LHC, of a novel hybrid model is presented in which the usual pre-hydrodynamic and viscous relativistic fluid dynamic (vRFD) stages are replaced by a {\it viscous anisotropic hydrodynamic} (VAH) core that smoothly interpolates between the initial expansion-dominated, approximately boost-invariant longitudinally free-streaming and the subsequent collision-dominated (3+1)-dimensional standard vRFD stages. This model yields meaningful constraints for the temperature-dependent specific shear and bulk viscosities, $(\eta/s)(T)$ and $(\zeta/s)(T)$, for temperatures up to about $700$\,MeV (i.e.\ over twice the range that could be explored with earlier models). With its best-fit model parameters the calibrated VAH model makes highly successful predictions for additional $p_T$-dependent observables for which high-quality experimental data are available that were not used for the model calibration.  
}
\maketitle
%

\noindent{\bf 1.\ Introduction.} Modeling ultra-relativistic heavy-ion collisions dynamically is challenging since the dense matter created in the collision passes through multiple stages of vastly different energy densities, requiring different theoretical frameworks built on different sets of dynamical degrees of freedom. State-of-the-art simulations employ so-called {\it hybrid models} consisting of several different evolution modules connected through sophisticated interfaces. Each module implements a different set of approximations to the underlying fundamental theory, to make it practically manageable. Typical hybrid models consist of an {\it initial conditions} module describing the deposition by the colliding nuclei of a part of their energy and conserved charges in the collision zone, followed by a {\it far-off-equilibrium pre-hydrodynamic} stage that evolves the resulting medium microscopically, either through interacting quantum fields or via kinetic transport theory for partonic phase-space distribution functions, whose output is fed into a {\it viscous relativistic fluid dynamics (vRFD)} module which propagates the emerging (never fully equilibrated) quark-gluon-plasma (QGP) fluid until it passes through the color-confining quark-hadron phase transition. Shortly after that one implements (on a 3-dimensional hypersurface inside the hadronic phase) a {\it particlization} interface that again changes language from macroscopic hydrodynamic to microscopic hadronic degrees of freedom, which are then evolved with a {\it hadronic afterburner} module by solving a set of coupled relativistic Boltzmann equations until the matter is so dilute that all strong interactions cease and all strong decays of unstable hadronic resonances have happened. Quantum statistical correlations between two or more final-state hadrons can be implemented on the afterburner output by hand. The final stable particles emitted from the model can be analyzed in exactly the same way as the experimentally observed particles emitted from a real heavy-ion collision, and the resulting observables compared. 

For each of these five modules multiple variations exist on the market, reflecting slightly or not-so-slightly different model assumptions. Considering all technically possible module combinations, literally hundreds of hybrid models for the fireball evolution can be contemplated. Each of these hybrid models has about 1--2 dozen model parameters, many of them representing fundamental intrinsic properties of the fireball medium that can not (yet) be calculated from first principles but whose most likely ({\it maximum a posteriori} or MAP) values and their uncertainties must be inferred from experimental observations by Bayesian model calibration. Since state-of-the-art hybrid models are highly complex and include many physical details in order to achieve a high degree of quantitative predictive accuracy, they are numerically very costly. Computing and exploring the posterior probability distribution of the model parameters in their high-dimensional parameter space is computationally infeasible without advanced acceleration techniques that replace the full hybrid model by computationally inexpensive surrogate models (i.e. {\it model emulators} for each observable), using advanced machine learning (ML) techniques. For a first idea of some of the ML techniques that have been developed for and used in recent hybrid model calibration campaigns we refer the reader to Refs.~\cite{Bernhard:2016tnd, Bernhard:2018hnz, Bernhard:2019bmu, JETSCAPE:2020shq, Nijs:2020ors, Nijs:2020roc, JETSCAPE:2020mzn, Liyanage:2023nds}, but the literature on this subject is growing daily \cite{Paquet:2023rfd}. In this contribution we compare model predictions from three variants of the JETSCAPE SIMS model \cite{JETSCAPE:2020shq, JETSCAPE:2020mzn} with those from  the VAH model \cite{Liyanage:2023nds} for which we had in-house access to all the necessary model output and model emulators.


\begin{SCfigure}
\centering
\includegraphics[width=0.695\textwidth,clip]{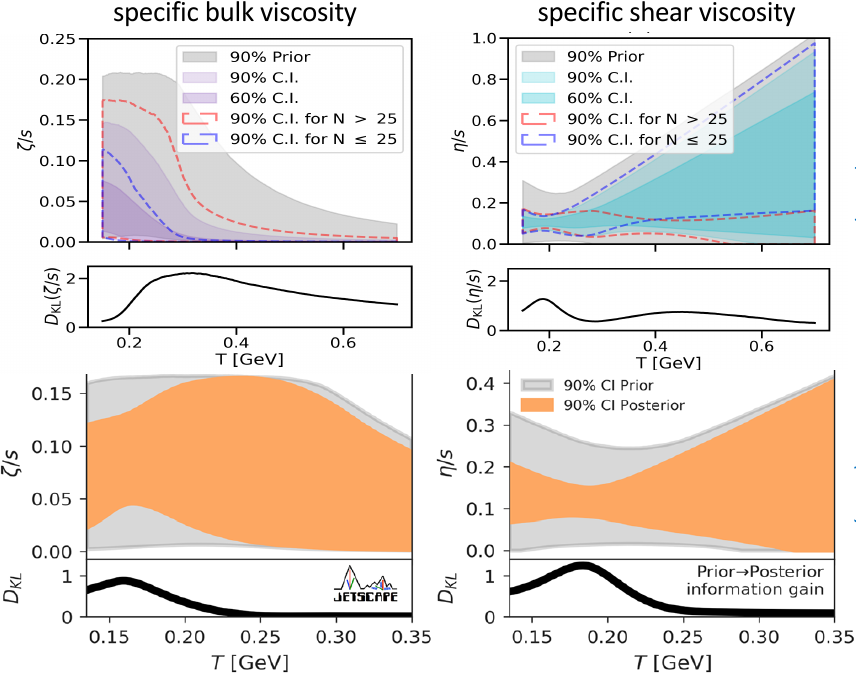}
\hspace*{1mm}
\caption{Prior (grey) and posterior (colored) credibility intervals for the temperature-dependent specific bulk (left) and shear (right) viscosities inferred from the VAH model \cite{Liyanage:2023nds} (top) and from the JETSCAPE SIMS model \cite{JETSCAPE:2020shq} (bottom). The narrow subpanels show the Kullback-Leibler relative entropy $D_\mathrm{KL}$ which measures the information gain provided by the experimental data used for the model calibration. Note the different temperature ranges covered by the top and bottom panels.}
\label{F1}       
\end{SCfigure}

\noindent{\bf 2.\ QGP viscosity constraints in the VAH model.} The JETSCAPE SIMS model \cite{JETSCAPE:2020mzn} uses T$_\mathrm{R}$ENTo initial conditions which are evolved by free-streaming (FS) before matching to vRFD after some time that is inferred from the data. 
Three particlization prescriptions (GRAD, CE, and PTB, see \cite{JETSCAPE:2020shq, JETSCAPE:2020mzn} for details) were studied for the particlization from vRFD to the SMASH afterburner. The VAH model \cite{Liyanage:2023nds} replaces the free-streaming and vRFD stages by a single {\it viscous anisotropic hydrodynamic} module, coupled to its own particlization routine, that smoothly interpolates between the initial expansion-dominated, approximately boost-invariant longitudinally free-streaming and the subsequent collision-dominated (3+1)-dimensional standard vRFD stages. This eliminates the free-streaming time parameter (and an associated discontinuity in the equation of state \cite{JETSCAPE:2020mzn}) and extends the sensitivity of the model to the QGP  viscosities into the far-off-equilibrium pre-vRFD stage, allowing for their inference from the data to much higher temperatures than in other available evolution models. This is shown in Fig.~\ref{F1} where the Kullback-Leibler divergence $D_\mathrm{KL}$ (quantifying the information gain provided by the experimental data) is seen to be non-zero up to temperatures of around 700\,MeV in the VAH model (even if the shear viscosity remains poorly constrained at the highest temperatures). At low temperatures the VAH model yields tighter constraints than JETSCAPE SIMS. The substitution of FS+vRFD by VAH in the VAH hybrid model relies on the observation in \cite{Jaiswal:2021uvv} that for Bjorken flow, which quite well approximates the full (3+1)-d flow pattern during the first 1--2\,fm/$c$ in Pb-Pb collisions at the LHC \cite{McNelis:2021zji}, VAH accurately reproduces the microscopic kinetic evolution of the full energy-momentum tensor at all times, for arbitrary initial conditions, for both massless and massive partonic gases.

\begin{figure}[h!]
\centering
\vspace*{-4mm}
\includegraphics[width=\textwidth,clip]{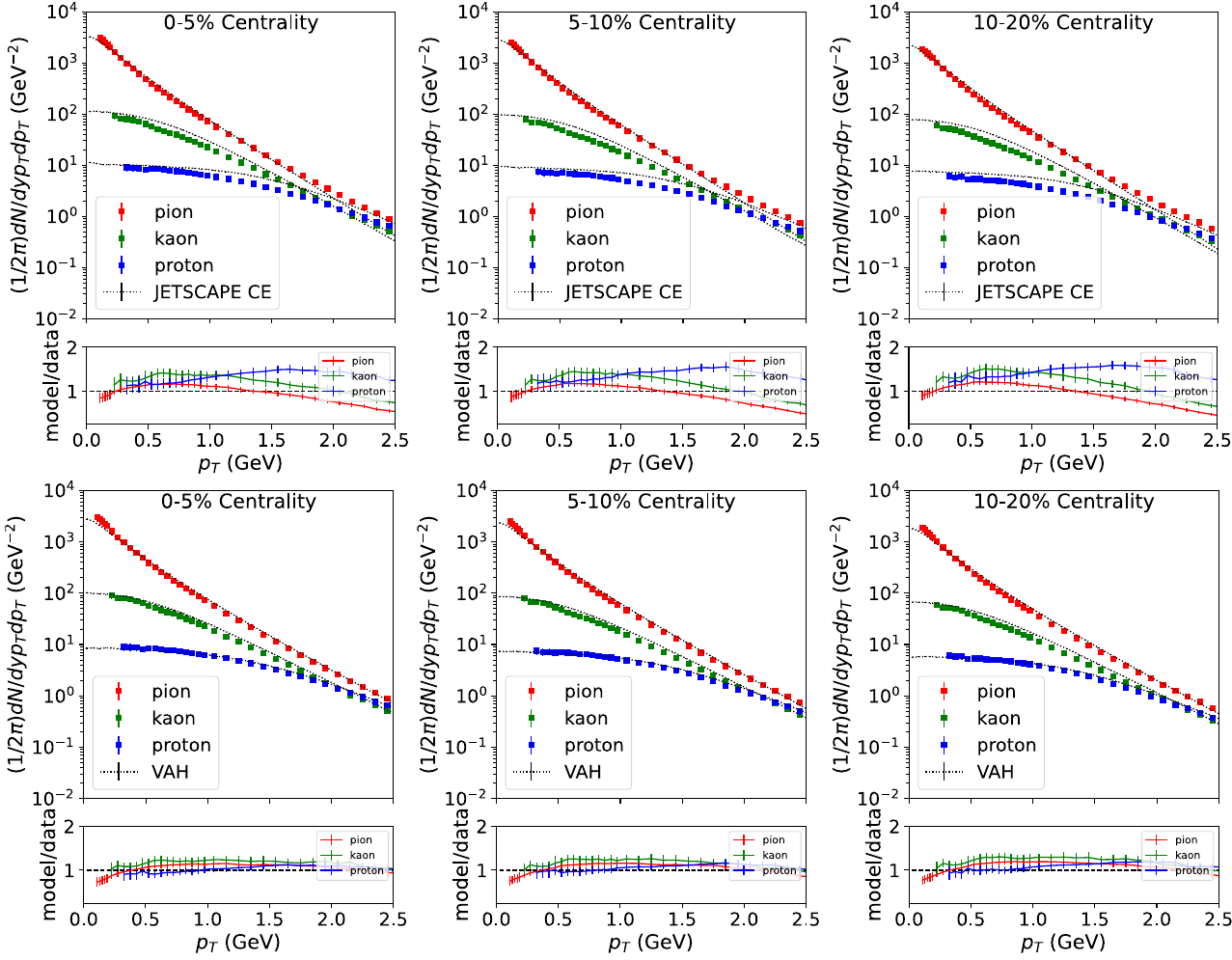}\\[-2ex]
\caption{Dashed lines show the $p_T$-spectra of identified hadrons ($\pi$ (red), $K$ (green), $p$ (blue symbols)) predicted by the best-fit (MAP) parameters of the calibrated JETSCAPE SIMS model with Chapman-Enskog (CE) particlization  \cite{JETSCAPE:2020shq,JETSCAPE:2020mzn} (top row) and the calibrated VAH model  \cite{Liyanage:2023nds} (bottom row), for 2.76\,$A$\,TeV Pb+Pb collisions at three collision centralities (left to right: 0-5\%, 5-10\%, 10-20\% \cite{ALICE:2015dtd}). The model/data ratios shown below each panel bunch significantly more tightly around 1 for the VAH model (bottom) than for the JETSCAPE model (top).}
\label{F2}       
\end{figure}

\begin{figure*}
\centering
\includegraphics[width=\textwidth,clip]{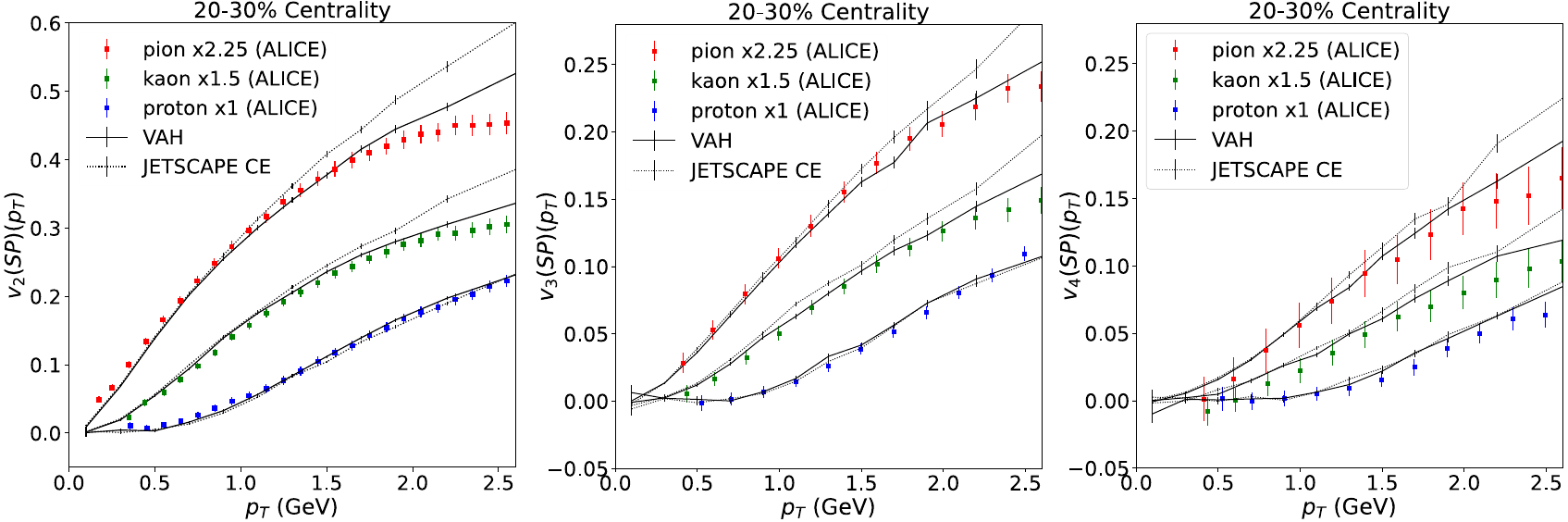}\\[-2ex]
\caption{Lines show for identified hadrons $\big(\pi$ (red, scaled by a factor 2.25 for visibility), $K$ (green, scaled by 1.5), and $p$ (blue)$\big)$ the $p_T$-differential elliptic ($v_2\{\mathrm{SP}\}(p_T)$, left), triangular ($v_3\{\mathrm{SP}\}(p_T)$, middle) and quadrangular ($v_4\{\mathrm{SP}\}(p_T)$, right) flows predicted by the best-fit (MAP) parameters of the calibrated JETSCAPE SIMS model with Chapman-Enskog (CE) particlization \cite{JETSCAPE:2020shq,JETSCAPE:2020mzn} (dotted lines) and the calibrated VAH model \cite{Liyanage:2023nds} (solid lines), for 2.76\,$A$\,TeV Pb+Pb collisions at 20-30\% centrality measured by the ALICE Collaboration \cite{Noferini:2012ps,ALICE:2011ab}. At this and at all other available collision centralities the VAH predictions agree better with the data than the JETSCAPE SIMS predictions, for all three particlization models studied in \cite{JETSCAPE:2020shq,JETSCAPE:2020mzn} (see \cite{Gantenberg_thesis}).}  
\label{F3}       
\vspace*{-5mm}
\end{figure*}

\noindent{\bf 3.\ Predictive power of the calibrated VAH and JETSCAPE SIMS models.} The VAH and JETSCAPE SIMS models were calibrated on $p_T$-integrated observables from Pb-Pb 
collisions at the LHC. Here we test the ability of both calibrated models to predict the most accurately measured $p_T$-differential observables: the $p_T$-spectra of $\pi$, $K$, $p$ in Fig.~\ref{F2} and their $p_T$-differential 
%
%
anisotropic flows $v_n(p_T)$, $n=2,3,4$, in Fig.~\ref{F3}. The consistent improvement of the description of the experimental data by VAH across collision centralities suggests that its more accurate modeling of the far-off-equilibrium early evolution stage is indeed important and pays dividends.

%
%

\noindent {\bf Acknowledgements:}  This work was supported by DOE (Office of Nuclear Physics), NSF (BAND and JETSCAPE Collaborations), and the Ohio State University Emeritus Academy.
%


\vspace*{-3mm}
\bibliography{biblio.bib}

\end{document}